# "My Zelda Cane": Strategies Used by Blind Players to Play Visual-Centric Digital Games

DAVID GONÇALVES, LASIGE, Faculdade de Ciências, Universidade de Lisboa, Portugal
MANUEL PIÇARRA, LASIGE, Faculdade de Ciências, Universidade de Lisboa, Portugal
PEDRO PAIS, LASIGE, Faculdade de Ciências, Universidade de Lisboa, Portugal
JOÃO GUERREIRO, LASIGE, Faculdade de Ciências, Universidade de Lisboa, Portugal
ANDRÉ RODRIGUES, LASIGE, Faculdade de Ciências, Universidade de Lisboa, Portugal

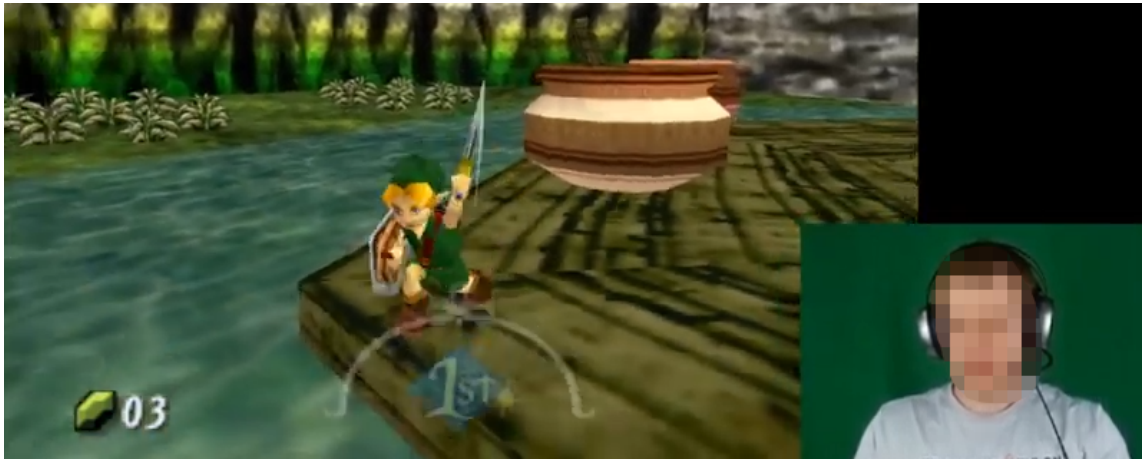

Fig. 1. Screenshot of a totally blind YouTuber (C4), playing The Legend of Zelda: Majora's Mask. C4 hits objects around by using the character's sword in order to feel the surroundings.

Mainstream games are typically designed around the visual experience, with behaviors and interactions highly dependent on vision. Despite this, blind people are playing mainstream games while dealing with and overcoming inaccessible content, often together with friends and audiences. In this work, we analyze over 70 hours of YouTube videos, where blind content-creators play visual-centric games. We point out the various strategies employed by players to overcome barriers that permeate mainstream games. We reflect on ways to enable and improve blind players' experience with these games, shedding light on the positive and negative consequences of apparently benign design choices. Our observations underline how game elements are appropriated for accessibility, the incidental consequences of audio design, and the trade-offs between accessibility, agency, and engagement.

## 1 INTRODUCTION

Accessible digital games for blind people consist mostly of games exclusively designed for their needs [2, 4, 18, 45, 52]. While these enable blind people to enjoy the medium, they drastically contrast with the complex virtual environments, behaviors, and content available in mainstream games [18, 41]. This paradigm has serious consequences, as it makes it hard for blind people to engage in mainstream gaming culture and limits their ability to share gaming experiences with others, particularly those who are not blind [18].

While most games are stereotypically designed for sighted play, it does not stop some blind players from experiencing mainstream titles, even when these are devoid of accessibility efforts. This is observed in online video-sharing platforms, where blind users broadcast and share their experiences with mainstream gaming. These videos enable us to identify the barriers and associated coping strategies, which can provide invaluable insights to inspire new accessibility practices, as has happened before in the industry [15].

Prior research has focused on recounting blind players' experiences with accessible (e.g., audio) games [45], or solely identifying barriers at surface level [2, 18, 38]. The analysis of online videos has been successfully used before to examine people with disabilities' practices in other contexts (e.g., cooking) and derive implications for the design of more inclusive technologies [3, 13, 28, 29, 50]. In this work, we leverage a similar methodology to understand how blind people are playing mainstream digital games, focusing on the strategies used to navigate, perceive, and interact with the virtual world. We sought to answer the following research questions (RQ):

- *RQ1:* What barriers do blind players face when playing mainstream digital games that mostly present visually-reliant challenges (in particular, navigation and interaction with virtual objects)?
- *RQ2:* What strategies do blind players leverage in order to work around inaccessible gameplay?

We conducted an online ethnographic study where we observed 33 playthroughs (consisting of multiple videos) and 32 showcase videos, uploaded by 14 blind users on YouTube, covering a total of 40 mainstream games and over 70 hours of video content. Our results show how blind players act on the environment to understand their in-game surroundings (e.g., by purposefully bumping into walls), leveraging landmarks to navigate. We describe the burden of preparation and keeping track of information not always available. We further highlight the new styles of gameplay that arise (e.g., co-pilot), as well as community efforts in creating tools to overcome gaps in accessibility. By providing an in-depth understanding of these experiences, we call forth opportunities to shape the design of digital games toward welcoming both sighted and blind players to the medium and promoting inclusion.

## 2 RELATED WORK

Integrating blind players' needs in mainstream games is still an open challenge, given the strong focus on high-end graphics and vision-dependent challenges (e.g., shooting). Consequently, a large number of individuals are excluded from, or have a diminished experience, playing mainstream titles [2, 18, 38]. In this section, we discuss 1) accessible games, and current accessibility practices in the industry; 2) research exploring blind-accessible gameplay; and 3) navigation in virtual environments.

### 2.1 Playing blind

Blind-accessible games are usually games specifically designed for this population, namely audio games, which present auditory challenges to players and typically do not have graphics at all [2, 45]. Text-based games are also generally accessible to blind people when compatible with screen reading software [2, 18, 38]. Some fighting games are also shown to be largely accessible to blind players [2], with the challenge consisting of perceiving the different combat sounds and reacting accordingly.

Recent efforts from the industry include the launch of new consoles from Microsoft and Sony, where accessibility is prioritized (e.g., screen reading and interface resizing since release). Microsoft has launched other products, such as the adaptive controller[1] and the co-pilot feature[2], which are shown to significantly improve the experiences of people with various disabilities. Accessibility guidelines for game design have been published and disseminated in different formats [7, 20, 25, 31]—in particular, a note for the Xbox Accessibility Guidelines [31], which include detailed guidance, extensive examples, and supplemental resources for effective implementation.

These efforts are also reflected in the actual design of recent games. The Last of Us Part II [36], released in 2020, features more than 60 accessibility options and is the first mainstream game to purposefully implement design solutions that allow blind players to complete the game. It provides options to navigate a complex three-dimensional environment, including navigation assistance and auditory representation of objects and enemies. One of the central accessibility features of the game was inspired by the way gamer (and YouTube content creator) Brandon Cole navigated in the game Resident Evil 6 [15].

While some games accessible to blind players include navigation, this is usually confined to two-dimensional grid surveying [34]. Previous works point out the design tension between oversimplifying the challenge and making it too overwhelming, given the low resolution of audio and haptic feedback [2, 18, 41]. It is not a trivial challenge to achieve the sorts of sophisticated interactions existing in mainstream games solely relying on audio and haptic feedback. Even in The Last of Us Part II, while most challenges are blind-accessible, others (including navigation) are partially or totally automated by accessibility features.

In reality, digital games are still largely inaccessible to blind players and accessible alternatives are not comparable to the variety and complexity of challenges presented in mainstream titles [2, 18, 41]. Research is needed to understand barriers elevated by current game design practices and find ways (and associated trade-offs) to bridge this gap.

### 2.2 Research on blind-accessible game design

Diverse ways of using audio feedback have been investigated toward more accessible experiences, both in two and three-dimensional environments [1, 5, 30, 34, 41, 43, 47]. For instance, in Audioquake [5], an adapted version of the popular first-person shooter Quake, researchers explore different rendering strategies to create accessible navigation: earcons i.e. artificial sounds, often following musical conventions to alert users, and auditory icons i.e. audio effects that sound like real-world objects and events. The authors highlight the advantages of each approach, in particular how earcons can provide information in reduced practical time and how auditory icons are more intuitively associated with the elements they depict.

Audio-based gameplay can also be inspired by real-life navigation systems, including compasses, sonar, and GPS [47]. In PowerUp [43], players are able to scan the surroundings for items of interest, lock objects, and automatically walk toward relevant positions. Similarly, NavStick [34] is an audio-based tool that enables blind players to probe their surroundings at their own pace, by scanning a specific direction at a time with the controller joystick. The tool showed potential to make three-dimensional navigation more accessible in mainstream gaming, still fostering a sense of agency. Research has also leveraged haptics for game navigation with custom devices, which provide novel ways to create accessible gameplay [30, 51].

---

[1]Xbox Adaptive Controller. https://www.xbox.com/en-US/accessories/controllers/xbox-adaptive-controller (Last visited on August 17th, 2022)
[2]Copilot on Xbox One. https://beta.support.xbox.com/help/account-profile/accessibility/copilot (Last visited on August 17th, 2022)

Inclusive multiplayer is a major challenge usually disregarded by accessibility efforts, both from industry and research [18, 19, 21]. Previous works [18, 38] highlight the difficulties of playing with others from the perspective of people with disabilities, with the feeling of unfairness being a primary reason for avoiding shared experiences. Universal game design [21] proposes strategies based on multimodality and adaptation, where people can play with the interface adequate to their needs (e.g., one-button interaction, text-to-speech), but has only been successfully applied in simple games (e.g., chess). Recent works propose an explicit asymmetry in the gameplay, not only in the interface and way of playing but also in the challenges presented to each player [16, 19].

## 2.3 Navigation in virtual environments for blind people

According to the literature, navigation is one of the major barriers for blind players to access mainstream games [18, 34]. Current solutions with virtual environments focus on supporting mobility training and mental mapping of real-world locations [12, 14, 23]. Design strategies based on real-world interactions, such as virtual canes [40, 53] and echolocation [1] were previously explored in this context. Curiously, research shows that less immersive, but practical solutions—jumping between decision points, instead of regular step-by-step walking—can also be useful and result in faithful mental representations of the real-world [23, 24]. This suggests that designers and developers may not need to replicate users' real-world techniques and behavior in order to enable them to successfully interact with virtual environments and therefore are encouraged to explore out-of-the-box solutions. However, for game worlds, we have yet to understand the trade-offs between practical, accessible, and immersive navigation solutions.

## 3 ONLINE ETHNOGRAPHY: BLIND PLAYERS' STRATEGIES TO PLAY VISUAL-CENTRIC GAMES

Online video-sharing platforms allow users worldwide to share their unique gaming experiences with online communities, which is common both in live-streamed and pre-recorded formats. Videos uploaded, including full playthroughs of a game, can provide a window to examine a variety of digital play experiences. Similar to previous work [3, 28, 29, 50], we conducted an online ethnographic study [35], where we observed blind users playing digital games on YouTube. As research has previously characterized blind players' experiences with audio games and other accessible alternatives [2, 45], we focus on their experiences with mainstream digital games, where challenges are mostly visual and involve some form of navigation and interaction with virtual objects. Understanding the barriers and, especially the strategies leveraged in these games, can pave the way for games that engage both sighted and visually impaired players alike, enabling more players to join in mainstream gaming culture. We identified a group of 14 YouTube channels, and observed 33 playthroughs (P1-P33) and 32 showcase videos (S1-S32), totaling over 70 hours of content.

### 3.1 Procedure

We aimed to understand the strategies used by blind players, by observing individual experiences and identifying accessibility barriers rooted in the gameplay (in particular, navigation and interaction with virtual objects), as well as the coping techniques leveraged. Our selection presents a variety of game genres (e.g., shooter, racing), spatial settings (e.g., first-person perspective, top-down), and interaction challenges (e.g., driving vehicles, combat). Below, we detail our data collection, searching and filtering procedures, and analysis. This process is summarized in Figure 2.

*3.1.1 Data collection.* We created search queries based on combining keywords related to vision impairment (i.e. "blind", "visually impaired", "sightless";) and gaming (i.e. "game", "games", "gamer", "gaming", "video gamer", "video game", "video games", "video gaming"). We searched Youtube[3], on March 10th, 2022 for videos that matched on the channel's name, video's title, description, and/or tags, which led to 10 453 results.

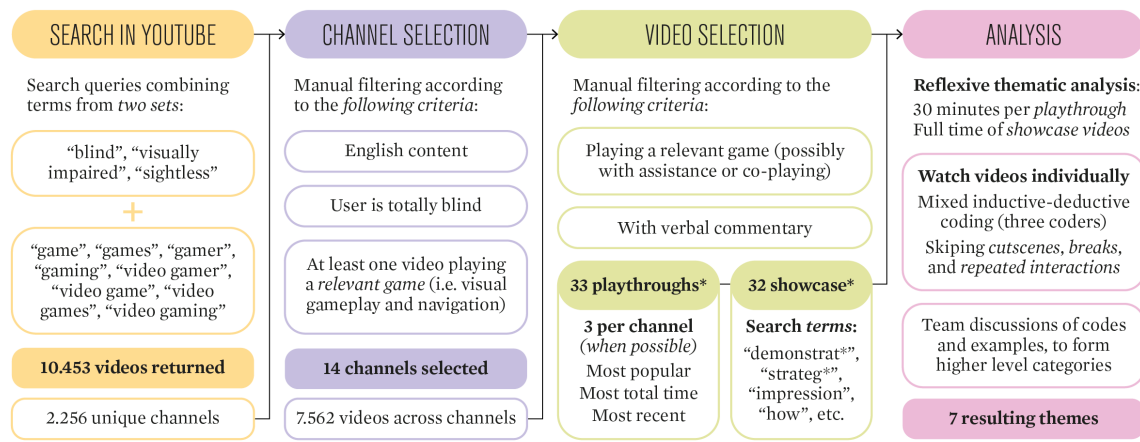

Fig. 2. Graphical summary of the study procedure.

*Channel selection:* We identified 2 256 unique channels across the video dataset. We manually checked the channel information and videos for the following inclusion criteria: 1) content is in English; 2) contains at least one video playing a game with visual gameplay, where navigation is part of the challenge; and 3) indication the user is totally blind (not able to leverage visual feedback); resulting in 14 YouTube channels (C1-C14) included.

*Video selection:* The 14 channels had over 7 500 videos uploaded. For each channel, we assembled one playthrough per game—a playthrough is a playlist containing all videos playing a single game—including only videos with verbal commentary, as it would enable us to have a greater understanding of the context of play.

We selected three playthroughs per channel to analyze, according to three different criteria: 1) playthrough with the most viewed first video; 2) playthrough with the most total time; and 3) playthrough most recently uploaded. We maximized game variety, by selecting first from channels with fewer unique games, and advancing in order to the channels with more variety, ensuring that no games were repeated, if possible.

We added individual videos we found to be especially useful, consisting of showcase videos i.e. concise videos where the user explains and demonstrates how they play a game (or a part of it). To collect these videos, we conducted a separate search on each channel, based on a set of keywords (e.g., "how", "introduction") and manually filtered out videos that were not in accordance with our definition of showcase video.

*Data analysis:* We conducted a reflexive thematic analysis of the final selection of videos. Our analysis protocol consisted in watching each video individually and annotating gameplay interactions and users' comments relevant to inform our research questions (i.e. barriers and workarounds encountered). For playthroughs, we started from the oldest upload date as it is more likely to showcase the origins of the strategies employed throughout the gameplay. We skipped non-player-driven gameplay footage (e.g., cutscenes, menu configuration, breaks) and when the user was repeatedly doing the same interactions. We used a two-minute window skip. Coders could always trace back if needed to understand the context of the current interaction being observed. We analyzed a total of 30-minutes of gameplay per playthrough (given that some were over 30 hours) and skips did not count toward this limit. For the individual showcase videos, we analyzed the videos in full.

---

[3]YouTube Search Python Library. https://github.com/alexmercerind/youtube-search-python

DG, MP, and PP were mostly involved in coding. They started by independently observing the same two playthroughs (Animal Crossing from C9 and God of War from C3) according to the protocol just described and independently annotated the videos with written observations and codes. They then met, first the three of them, and then with the whole team, to discuss and consolidate the procedure (i.e. skipping and annotation details). DG and MP proceeded to annotate the playthroughs, dividing them evenly. PP was responsible for coding the showcase videos. All other researchers familiarized themselves with a selection of representative videos chosen by the three coders.

We followed the steps of a reflexive thematic analysis outlined by Braun and Clarke [8] and supported by a workable example [10]. The coding was mostly inductive but deductively framed, given we were focused on annotating barriers and strategies to overcome them. The whole team had five sessions of about three hours each where the codes, observations, and quotes were discussed, videos were revisited, and bodystorming was enacted to facilitate the comprehension of barriers and strategies. These sessions served to explore multiple assumptions and interpretations of the data within the research team, with the goal of collaboratively sense-checking and identifying latent constructs across the data. During the sessions, one researcher was collating the discussions on a whiteboard creating groupings, higher-level categories, and relationships between the concepts discussed. Throughout these meetings, we formed and named our themes, which we present below.

The researchers involved are all sighted. DG, MP, and PP are junior researchers in their 20s. JG and AR are senior researchers in their 30s, working on the topic of accessible technology for approximately a decade. All authors regularly play digital and tabletop games and watch gaming-related videos on YouTube, except JG who plays occasionally.

### 3.2 Ethical considerations

The content creators were contacted and informed about the study, its goals, and which videos we selected from their channel. In these contacts, we made it clear we will send the final version of the article, if and once it gets published. The content creators were also asked for authorization to use the screenshots included in the paper (C4 and C9) and to include their channel's name in the supplementary material (all). 9 of them responded, expressing enthusiasm for the project and giving their consent. The other 5 did not respond and, as such, remain anonymous. Although this prevented us from providing the full list of channels in the supplementary material, we describe the data collection procedure in detail to maximize study replicability.

## 4 FINDINGS

Our collection includes videos covering 40 games. Some of these are played in a two-dimensional[4] top-down perspective (6), while others are played in a three-dimensional first-person (14) or third-person (20) perspective. In most videos, blind players are playing on their own, but in others, they are playing with other people, either in multiplayer, sharing control of the game through co-pilot features, or engaging with spectators (details in section 4.7). In 12 of the playthroughs, it is explicit the user is playing the game for the first time while in the remaining (and showcase videos), it is observable the user is acquainted and had past experiences with the game. Detailed information regarding the videos watched, including settings and external tools, as well as games observed is available online[5]. Additionally, the games observed are listed in Table 1.

---

[4]Three of them present multi-height environments, but navigation is controlled in two axes
[5]Detailed information about the final video collection and games. https://osf.io/hpme2/?view_only=e861d2c997ff43ccba23587ca149c824

Table 1. Basic information about the games observed: Name of the game or franchise (we opt to abbreviate some names throughout the paper) and navigation type: two-dimensional top-down (2D), three-dimensional third-person (TP), or first-person (FP). Some games were observed in more than one channel. Seven titles from CoD and two from Pokémon are included, totaling 40 different games from 31 different franchises (details in the full table provided).

| GAME / FRANCHISE | NV | GAME / FRANCHISE | NV | GAME / FRANCHISE | NV |
|---|---|---|---|---|---|
| **Animal Crossing** | 2D | **Call of Duty** (various) [**CoD**] | FP | Dungeons & Dragons [**Neverwinter**] | TP |
| [**Dark Souls**] III | TP | [**Dead Space**] 2 | TP | [**Destiny**] 2 | FP |
| [**Diablo**] III | 2D | [**Dragon Quest**] XI | TP | Elder Scrolls V: [**Skyrim**] | FP |
| [**Final Fantasy**] VII Remake | TP | [**Forza**] Horizon 5 | TP | [**Gears**] of War 4 | TP |
| [**Gears**] 5 | TP | **God of War** | TP | Grand Theft Auto V [**GTA**] | TP |
| [**Halo**] Infinite | FP | [**Horizon**]: Zero Dawn | TP | **Kingdom Hearts** | TP |
| Left 4 Dead 2 [**L4D2**] | FP | **Minecraft** | FP | Minecraft Dungeons [**MC Dungeons**] | 2D |
| **Pokémon** (various) | 2D | **Portal** | FP | [**Rainbow**] 6: Extraction | FP |
| [**Ratchet & Clank**]: Rift Apart | TP | [**Resident Evil**] 6 | TP | [**Star Wars**]: Battlefront 2 | TP |
| [**Stardew**] Valley | 2D | The Last of Us Part II [**TLoU**] | TP | The Legend of [**Zelda**]: Majora's Mask | TP |
| The [**Matrix**] Awakens | TP | **Warframe** | TP | World of Warcraft [**WoW**] | TP |

The games observed present different levels of accessibility but—with two exceptions where users try the game for the first time and quickly give up (C2 playing Rainbow and C14 playing Animal Crossing)—the videos show the users involved in the gameplay, surpassing the different challenges posed. As expected, many of these challenges are heavily dependent on vision, which affects players' performance and derails player experience from what was intended by design—e.g., taking almost five minutes to exit a room through the door. To overcome these barriers, players often rely on persistent trial-and-error and ingenious strategies that leverage the game's own mechanics and external tools. In some cases, users resort to other online videos (e.g., walkthroughs) and sighted assistance.

Below we present our themes (summary in Table 2), focusing on describing the strategies created and adopted by these players, as well as contextualizing the barriers imposed by the design of these games. We illustrate our findings with descriptions of gameplay interactions and quotes extracted from players' commentary.

### 4.1 Understanding the surroundings

All videos collected involve navigation challenges where the player is required to move the avatar from one point to another, either in a two- or three-dimensional space. The perception of the virtual environment is essential to progress. This includes perceiving the boundaries and obstacles as well as interactable objects and entities.

*4.1.1 Grasping the soundscape.* Certain games deliver rich soundscapes to the players, allowing them to extract essential information about the environment and the objects in it. It is common to observe players paying close attention to the different sounds and trying to map them to what they represent. In showcase videos, users center their discourse on explaining how they can identify different game elements and interactions by distinguishing the sounds they emit: "*Teleportation does have a distinct sound, so I know every time he teleports*" (C2, playing *Kingdom Hearts*). For most games observed (36 out of 40), blind players make use of **spatial audio** to perceive the position and direction of these elements. Usually, players can also distinguish different areas of the environment thanks to how the **soundscape changes**. Sound treatment (e.g., distortion, pitch) is useful in many cases—for instance, discerning when the character enters a building, as the sound becomes muffled. In most games (e.g., *Animal Crossing*, *Skyrim*), **footstep sounds** are different when walking on different terrains (e.g., sand, grass, stone).

Table 2. Summary of findings, organized into the resulting themes (T) and respective untackled barriers.

| | |
|---|---|
| **T1. Understanding the surroundings.** Leveraging spatial audio, discerning sound effects (e.g., footsteps, voice lines) and soundscape changes (e.g., audio treatment). Feeling through bumping and interacting with objects. | *Surroundings (Untackled barriers):* Elements not interacting with the player are often silent; Time-sensitive challenges hinder feeling around; Occlusions. |
| **T2. Wayfinding in virtual environments.** Navigating based on landmarks (sound or collision) and authoring new ones; Re-orienting by reaching a familiar spot (respawning, save states); Semi-automatic navigation. | *Wayfinding (Untackled barriers):* Objective indications are visual-only (markers, text); Own movement is not perceived; Irrelevant sounds attracting players. |
| **T3. Dealing with perspective.** Remapping camera control, recentering through keybinds, and adjusting sensitivity; Leveraging aim assistance; Perceiving height changes based on landing sounds. | *Perspective (Untackled barriers):* Misunderstanding camera yaw (aiming too high or too low); Height changes are ignored and platforming is inaccessible. |
| **T4. Interacting with the world.** Experimenting with controls; Button mash to check for interactions; Avoiding fine-grained interactions (resorting to area effects); Curating abilities and features (accessibility paths). | *Interacting (Untackled barriers):* Prompts are inaccessible or do not provide context; Precise aligning and aiming; Complex interactions (e.g., stealth, taking cover). |
| **T5. Preparation, demand & cognitive load.** Memorizing controls; Maintaining a mental map; Consulting walkthroughs and guides; Unintuitive and overwhelming sounds; Keeping up with the game state (e.g., health). | |
| **T6. Automation & difficulty.** Settings automating or reducing the challenge; Playing a game differently but able to participate. | |
| **T7. Playing with others.** Sighted co-players and spectators describing the surroundings, menus, and controls; Co-piloting by distributing controls; Collaborating and gaining autonomy; Latency and cumbersome assistance. | |

Sounds are not always intuitively associated with what they represent. For example, when playing *MC Dungeons*, C5 interprets the sound of zombies approaching as "*wolves*" and the sound of showing the mini-map as crouching with the character. In Star Wars, C9 explains it is impossible to distinguish between soldiers and turrets shooting, given the same sound effect is used for both. Sometimes, the soundscape may consist of an overwhelming amount of different sounds playing simultaneously, including ambient sound (e.g., raining making it harder for C4 to navigate in *Zelda*), music, and voice acting. As such, blind players make use of audio settings to adjust the volume of each channel to their preferences, when possible.

*Perceiving interactions and entities.* Elements moving or directly interacting with the player are generally conveyed through audio across the games observed. For instance, the presence and position of non-playable characters (NPCs) can often be perceived thanks to their **footsteps**, **movement trails**, and/or **voice lines** (e.g., *Animal Crossing*). Still, in some games, these characters are totally **silent** and mostly ignored by blind players (e.g., *Stardew*).

Enemies' position and general behavior are discernible in most games during combat segments, thanks to the sound effects for moving and attacking (e.g., a monster growling). However, before engaging in combat, the presence of enemies is often only conveyed through visuals. In Horizon, C3 is able to discern robotic enemies before engaging in combat, by their "*walking patterns*". However, human enemies are silent until they start attacking, which—alongside the absence of a "*tone change*" for when the player is within the enemies' field of vision—hinders a stealthy approach.

While moving elements and interactions are often conveyed through audio, static elements rarely emit sound. Exceptions include elements whose basic behavior is associated with persistent sounds, such as water falling, fire, or a light flickering. These elements are usually essential for blind players to orient themselves in the environment (details in section 4.2.1). Voice lines delivered by characters in the game can also be helpful. In *Destiny*, the sidekick character informs there is a "*chest*" in the room, making C1 aware of it. Apart from these exceptions, users resort to specific strategies (e.g., bumping) to perceive silent elements such as boundaries, obstacles, as well as entities and items that are not directly interacting with the character.

*4.1.2 Bumping to feel the surroundings.* When first interacting with the environment, players' behavior is often erratic, advancing by walking in different directions and running into obstacles. Blind players are often able to tell when the character collides with an obstacle, as the **footstep sounds** change or stop completely and/or the game provides a **'bumping' sound**. The absence of appropriate feedback, on the other hand, hampers players' ability to perceive collisions: "*It doesn't stop running when you run into a wall, that is problematic*" (C2, playing *Final Fantasy*). Players often seek to **purposefully collide** with obstacles in order to feel the environment layout. For example, C9 enters the house in *Animal Crossing*, by walking around it in order to feel its collision box and find the front door. In *Destiny*, C1 navigates through a spiral staircase and ascertains that, by forming a mental map through bumping.

*4.1.3 Feeling through action.* In addition to bumping, blind players leverage various game actions to feel their surroundings. Some actions were built with that purpose in mind, such as the item scan in *Halo* and *TLoU*. It consists of sending a 'pulse' that propagates from the player's position, detecting interactable objects and entities around, and emitting a spatial sound to convey their position. Still, other actions are leveraged in unexpected ways. In *Animal Crossing*, C9 has to interact with (i.e. shake) the trees to discern the different sounds they emit when shaken. In *Zelda*, C4 sporadically uses the basic attack with the sword—to which they calls "**my Zelda cane**"—to hit obstacles and feel their shape and material (e.g., wood, brick) [Figure 1]. This is observed in other playthroughs as well (e.g., C1 and C6 playing CoD). In *CoD*, C1 shoots occasionally to ascertain if there is an obstacle up front. This player also uses pinging (i.e. a feature that allows players to designate specific locations with their aim, creating visual markers used to communicate with their team) to identify objects, as it triggers discernible voice lines (e.g., the character says "*gun here*" after pinging a weapon).

*4.1.4 Awareness during combat.* The aforementioned strategies require time and some degree of quietness to be leveraged. In time-sensitive sections, especially combat sections, perceiving the surroundings becomes impractical— players are usually overloaded with combat-related sounds, needing to quickly defend, counter-attack, or hide. Some elements are not audible at all and cannot be bumped with (e.g., visual markers to symbolize where projectiles will land). Perceiving **occlusions** becomes a prominent barrier, with players targeting enemies' direction, but running into and shooting walls, making them vulnerable to other attacking enemies. Curiously, in one video, C1, when playing *CoD*, explains they can tell when enemies are behind a wall, thanks to how the sounds they emit are "*muffled*". However, during combat we observed that even C1 often runs into and shoots walls, trying to defeat enemies behind them.

*Voice lines for accessibility.* In combat sequences, voice lines are occasionally used to provide cues for players, particularly when enemies are out of sight (e.g., "*look out, behind you!*" in *God of War*). With this information, blind players are able to successfully turn around and counterattack. It also provides the player a sense of priority, allowing them to ignore other sounds and focus on the closest enemy. Such cues can be further explored within the context of accessibility as they could be beneficial beyond just announcing incoming attacks from enemies.

## 4.2 Wayfinding in virtual environments

For players to successfully navigate through the game world, it is not only required they understand what surrounds them, but also where they are and should be heading, and how they can reach their objectives.

*4.2.1 Perceiving objectives and progress.* In mainstream games, objectives are usually communicated to the players through **message prompts** (e.g., '*new objective: follow the car*') and **visual markers**, as well as subtle visual hints, map design, and lighting [37]. In most games observed, these elements are solely conveyed through visuals, resulting in players uncertain of how to progress—or if they are progressing at all, as the outcomes of their actions are not always perceptible. For instance, when playing *Zelda*, C4 tries to turn left but the camera fails for some reason. Although visually it is clear the avatar did not turn, C4 assumes it did, ending up stuck and seemingly confused. It was also common to observe players accidentally backtracking, having no way to tell if they were heading in the correct direction.

*4.2.2 Leveraging and authoring landmarks.* Blind players often navigate by following a mental map of the scene based on specific elements present in the environment. Usually, these **landmarks** consist of elements that emit a **persistent spatial sound** (e.g., fire pit) or emit it on **collision** (e.g., bush). This helps players have a sense of position and orientation relative to sound (e.g., C4 perceiving an entrance in *Zelda* thanks to a torch on each side of it). Conversely, in some cases, sound landmarks can lure players to irrelevant locations or even away from the objective. In *Destiny*, C1 is facing the objective but turns left to follow the sound of water falling (ambiance, but prominent sound), making the player stray from the path.

In one exceptional case, C4 navigates by using the information on the absolute position of landmarks. In *Stardew* (with mods), C4 presses constantly a key to know the avatar's **coordinates**. By knowing the coordinates of every point of interest (e.g., the house, the shop), C4 is able to locate objects and navigate efficiently in the game.

*Mobile landmarks.* Moving objects and entities can also serve as landmarks for navigation. It is common for players to follow along with NPCs, by orienting themselves with the sounds they emit when moving (e.g., footsteps). In *Star Wars*, C9 follows the footsteps of their AI-controlled allied troops to navigate through the map. In games where NPCs are programmed to advance toward the objective, they are also leveraged by players to find their way (e.g., following the Ghost in *Destiny*'s introduction scene). Enemies in the path can also serve as a point of reference, given they usually emit sound. In a section of *Zelda*, C4 defeats all enemies, except for one, purposefully leaving it behind to serve as a landmark: "*I want to use it later as a signal to be able to get back into this area*".

*Authoring landmarks.* Some players re-appropriate the game mechanics to create new landmarks in the environment. For instance, in *Portal*, C4 creates new portals only to help orientation (portals emit a persisting sound). In *Stardew*, C4 relocates two decorative objects in the house as landmarks to quickly find the door and bed [Figure 3]. In *Animal Crossing*, C9 creates new objects in the environment to landmark locations (e.g., torches on each side of an entrance) [Figure 3]. This player highlights how this strategy enables an accessible environment, that looks natural for blind and sighted players alike (e.g., using bushes to mark and surround a patio, as they fit the context): "*Look, this is how you can make a game accessible and still make it look natural, like... like it's built into the game, you know? It doesn't have to take away from the game at all*".

*4.2.3 Getting lost & reorientation.* While navigating, it is common for players to get stuck in the same location for minutes on end, either aware or unaware, depending on the game feedback. This can lead to players not feeling confident to explore: "*I don't know if I want to dismount, because I might get lost*" (C7, playing *TLoU*). When players were uncertain of their whereabouts, they employed strategies to reorient themselves, mainly reaching for a familiar location. In *Star Wars*, C9 uses the possibility of **respawning** (through the menu). In both *Portal* and *Zelda*, C4 relies extensively on **quick saves and loads** (by pressing a key) to recover when getting stuck, lost, or straying from the right path, sometimes to avoid time losses during a timed challenge.

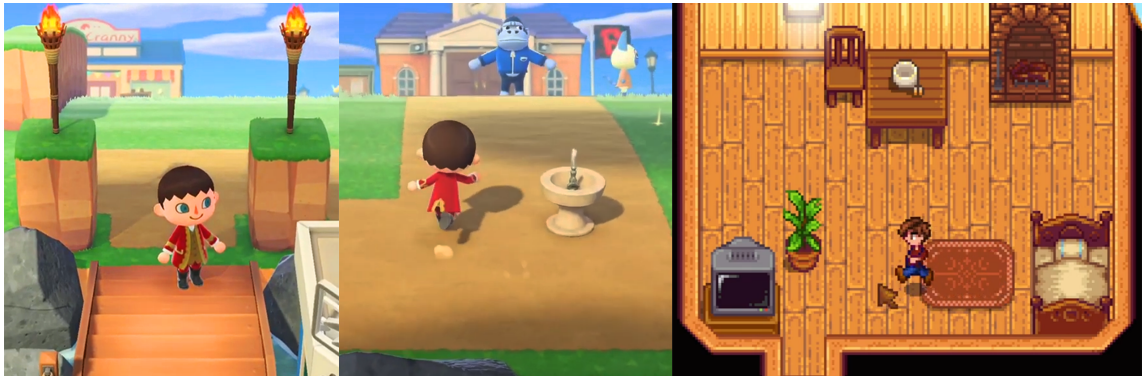

Fig. 3. Examples of how players leverage and author landmarks in the environment. In Animal Crossing (left), C9 uses torches to mark entrances and fountains to mark intersections; in Stardew (right), C4 uses the pot and the TV to align the avatar.

*4.2.4 Navigation aids.* Some games (5 out of the 40 observed) offer players navigation assistance, including **semi-automatic** forms of navigation. In *TLoU*, there is a feature to automatically align the camera with the next objective or focused interactable in their path. In *Forza*, an option lets the game automatically steer the car for the player. These accessibility features bypass the players' need to orient the avatar in the right direction. In *Pokémon*, C12 navigates by using a **pathfinding** mod that provides turn-by-turn instructions to reach any location in the environment (e.g., "*five steps up, two steps right*").

### 4.3 Dealing with perspective

In two-dimensional games, players only move on two axes, which made the game easier to navigate for some players: "*It's a top-down game, on your controller up is up, left is left, right is... like it's full 360 degrees. So it's pretty easy to play*" (C1, playing *Diablo*). However, in most games observed (34 out of 40), players navigate in three dimensions, requiring **height perception**. It was common to observe players unaware of height changes in the environment, struggling with gaps, jumps, and understanding targets' height. Also, the camera is controlled by the player, which determines the avatar's orientation (egocentric navigation), bringing forth new challenges.

*4.3.1 Camera yaw.* In three-dimensional environments, players were generally capable of pinpointing the horizontal direction of elements, thanks to spatial audio. However, they often struggled to understand and manipulate camera yaw, which is to be expected given the difficulties in conveying verticality through spatial sound [22, 46]: "*The main problem that I have with Left 4 Dead, that makes it a lot more difficult is aiming vertically. Like... when I move left or right, the sounds rotate around me, but when I move up or down, they don't*" (C9).

Players' awareness of this issue led them to reduce camera **sensitivity**, **recenter** the camera, or limit its movement: "*I also have everything mapped to the keyboard, so I don't have to use the mouse, so if I'm moving left and right, I don't end up looking up at the ceiling*" (C4, playing *Portal*). To recenter, players use specific key binds; interactions (e.g., taking cover, running); and as a last resort, interaction patterns created through trial and error: "*One thing I do in CoD is actually meleeing the floor [...] then I look up a little bit the aim assist [...] to pull it back up*"(C1).

**Aim assistance** can help but it comes with its own shortcomings. In most games the aim has to be close enough to 'lock on' but blind players' cameras are often nowhere near (e.g., aiming at the sky, against a wall). Aim assistance can also be the reason for decentering the camera, causing other issues (e.g., targeting uphill enemies resulting in the player unawarely aiming too high for other enemies).

## 4.4 Interacting with the world

When first playing a game, players have to slowly build a mental model of all the interactions and respective sounds. In *TLoU*, players are slowly introduced to these sounds, but in most other games players have to first experiment with controls and identify interactions based on the sounds they produce: "*Is that to block? That's what it sounds like*" (C1, playing *Dark Souls*). Players often express pleasure in discovering abilities and items that allow for new interactions and speak enthusiastically about the sound effects they emit. C1 expresses joy in trying different weapons in *CoD*, while C9 decorates the house in *Animal Crossing* with objects that trigger pleasant sound effects (e.g., a guitar).

*4.4.1 Interaction prompts.* During gameplay, players are prompted to do specific interactions at specific moments (e.g., open a door) and these prompts are usually only conveyed through visuals. To cope, players repeatedly press the input to check for available actions: "*There are no audio indicators for when your cursor highlights something in Skyrim, or Diablo or anything, so like, you have to basically button mash when you can't see*" (C1). In *TLoU*, where auditory cues are provided, players knew they had to interact but often not why. For instance, C10 got close to a car and received a cue to climb. The player then describes it as a "*jumping section*", climbing it as prompted, despite the car being just an element of the environment unrelated to progress.

*4.4.2 Fine-grained interactions.* Players often struggled with gameplay mechanics that required them to **precisely align** elements. In *Stardew*, we observed C4 failing to interact with a NPC, unable to face it (despite being right by its side), and subsequently assuming the character was not there. On the other hand, when players know the element is there but fail to interact, the experience becomes frustrating: "*I hate opening small chests. They seem to take me forever*"; "*You know, if you guys have watched my videos before, me finding doors is always fun*" (both quotes from C4, playing Zelda). In *TLoU* despite players being prompted to interact, they still misaligned and failed to interact repeatedly.

In combat-centered games, players tend to use attacks that have a large area of effect, avoiding precise combat (e.g., shooting). For example, in *Gears*, C3 either attacks enemies by using melee or the shotgun (which covers a large area when shot). In a particular level, C3 gives up the challenge after realizing that the enemies fly and need to be shot at a distance. In *Horizon*, instead of shooting enemies at a distance, C3 opts to hide in tall grass, whistle to attract enemies, wait for them to approach, and then do a stealth attack that can be triggered when the enemy gets close enough.

*4.4.3 Accessibility builds and paths.* In some videos, especially showcase videos, it is common for more experienced players to give advice and tips on how to take advantage of customization offered by the game. This involves the curation of specific **features** and **abilities** that can particularly benefit a blind player. For example, in *CoD*, C1 advises using the sixth-sense perk, which emits a sound effect when an enemy is close by. In *Gears*, C3 customizes the character's abilities to minimize taken damage and increase damage dealt by melee attacks. In some games, players strive to unlock specific abilities as early as possible to improve the experience from the start. In particular, in *WoW*, C11 mentions a mount that enables quick access to NPCs, instead of having to look for them. It is also common for players to suggest playing on specific maps (e.g., open maps, with fewer obstacles) and with specific characters (e.g., close combat characters). In most cases, these choices are essential for players to be able to play.

## 4.5 Preparation, demand & cognitive load

Information in digital games is often not accessible at all, let alone accessible at all times, which means players have to **memorize** and **keep track** of what happens in the game. Players often depend on their memory to be aware of the environment layout, interactions that happen during gameplay, and controls. Additionally, players often have to use and **configure tools** outside the game to be able to play, which is an extra effort: "*I feel like you need like eight hands to be a completely blind gamer*" (C14, playing *Animal Crossing* by holding the smartphone to read on-screen text while using the controller to progress).

*4.5.1 Walkthroughs & prior knowledge.* Blind players are usually deprived of the smooth learning curve, as intended by design in most games. As mentioned before, "learnability patterns" are mostly visual, including hints given through the use of color, graphic overlays, and lighting [37]. Contextual tooltips, designed to gradually introduce mechanics (e.g., "Press L1 to throw a grenade"), were not accessible in any game except *TLoU*. It is common for blind players to look for guidance by watching other **online videos** and reading **guides** that explain how to play, menu layouts, and walk the player through the game. For some games, consulting these kinds of materials is essential for players to be able to play. For instance, in *Pokémon*, C9 makes use of walkthroughs to navigate through story moments of the game, given these are not accessible. In some videos, players are able to navigate by having memorized descriptions and instructions on how to navigate in the environment.

*4.5.2 Maintaining a mental map.* Akin to sighted players, through repetition, players create and maintain a **mental model** of the environment: "*It takes a lot of repetition and practice but eventually you can memorize the map purely by sound*" (C1, playing *CoD*). In particular, C4 shows an extraordinary capacity to memorize the layout of games. In *Portal*, C4 is able to successfully go through the first levels independently, by having the puzzles and their solution memorized. In *Zelda*, C4 has all the levels mentally mapped out, describing what is on screen and progressing by employing established strategies for each level. In *Stardew*, C4 explored the map initially and memorized the coordinates to reach relevant locations. In this game, the player also memorized the behaviors of NPCs (the locations they are programmed to be, at different times of the day), in order to interact with them (given they are totally silent).

*4.5.3 Bandwidth conundrum.* In mainstream games, the amount of information present at each moment of the gameplay is catered to sighted play by design. For instance, during combat, players are often surrounded by groups of enemies, while having to dodge or block multiple incoming attacks and/or projectiles. Due to the **limitations of audio feedback**, players struggle to accurately pinpoint enemies: "*There's so much going on and you can't actually pick out where you're meant to be aiming, even in terms of melee, it's kind of difficult*" (C3, playing *Gears*). To cope, players would escape combat zones, pause, or even choose certain abilities to buy time to assess and strategize (e.g., C9 choosing a specific soldier with a shield in *Star Wars*).

In some cases, players are simply overwhelmed with different audio sources and it is difficult to discern the relevant (or priority) sounds (e.g., the sound of a generator muffling the sounds coming from Zombies in *CoD*). An overload may also emerge from having a big variety of sounds to indicate different things. Notably, in *TLoU*, there is a panoply of audio cues that correspond to different meanings (e.g., prompt to crouch, to jump over). While a full glossary is available to consult, it is cognitively demanding to recognize a sound that has no intuitive association with its meaning: "*There's a lot to get used to. A lot of sounds to get used to, a lot of, you know... identations, noises, audio cues, sound cues*" (C6).

*Keeping up to date with the game state.* Players have to mentally keep track of various information, due to **inaccessible interfaces** (e.g., health displays, inventory). It was common to observe players trying to shoot when ammunition was out or trying to heal when in full health. In some games, the game state is narrated whenever the player presses a certain input (e.g., *TLoU*). In others, specific cues (e.g., vibration when hit) and voice lines (e.g., "*no ammo*") can help. In *CoD*, C1 estimates how much money is available by dropping it in fixed quantities and adding up until there is none left.

### 4.6 Automation & difficulty

It was common for players to customize the settings (accessibility, difficulty, and others) to **reduce the challenge**. From making Zombies less aware of the avatar's presence and going invisible when crouching (*TLoU*), to even being invincible (*Ratchet & Clank*). Frequently, the explicit accessibility features reduce the challenge by automating part of the experience (e.g., auto-steer in *Forza*). While we are unsure of its direct effects on player experience, the reality is that, currently, without many of these features, playing vision-centric games would be virtually impossible. Many of these features are not always branded as accessibility and are part of the core experience, being therefore leveraged by many players (e.g., aim-assist with controllers).

*4.6.1 Playing your own game.* At times, players engaged in games in unconventional ways, recognizing how they have different expectations of the experience from sighted peers. When playing *GTA V*, C13 does not attempt to progress in any way, walking and shooting erratically, attempting to get into cars, and dying frequently. The player emphasizes playing their own way: "*So that's how I play video games. Basically I just... yeah... basically I just play, until it's just... I get bored*". This example somewhat maximizes player agency but, at the same time, the game inaccessibility shapes the experience. In another example, C10 plays *Forza* with auto-steer only using one button to accelerate: "*It's still fun [...] I'm still like... you know, like, actually feeling part of the game*". C10 commends the sound design and how it provides an immersive experience, despite the reduced control over the gameplay: "*that train passing by, I can hear it!*". Lastly, in *Animal Crossing*, C9 finds clams, whose position is only conveyed through visuals, by repeatedly and randomly digging the ground. C9 acknowledges the extra demand, but how it enables participation: "*It looks tedious but the fact that I'm able to do something... it's simple but I can't describe it other than normal. Because whenever a new game comes out, everyone's like 'oh wow, it's so awesome' and I can't usually play it. But like... day one I'm able to play this game [...] That's an experience I don't get to have very often*".

### 4.7 Playing with others

In 25 of the playthroughs, blind players were playing with others: with strangers (6) and/or friends (3) in multiplayer mode; with friends sharing control with co-pilot (6), only spectating (4) and/or a live audience (9). Additionally, in 4 videos, blind players were assisted by friends at one point during the video, by handing over the controller (2) or asking for a description of the gameplay (1). Below we discuss the various strategies employed to cooperate.

*4.7.1 Collaborating with players.* We observed blind players playing with acquaintances, sharing an experience through dialogue and, planning division of gameplay tasks. When playing with their regular friends, the collaboration evolved seamlessly with sighted players stepping in with additional cues and information about menus, the surroundings, and interactions: "*Am I facing fine?*" (C9, playing L4D2); [a co-player stops shooting to check on C9] "*Look left... yeah, you're fine*". Players also agree on the options that provide the most accessible experience: "*The map is called The Bridge, that one is easiest for me to play* (C9, playing L4D2)". When playing *Halo* against other blind people, C10 creates a custom match where only melee weapons are allowed, as aiming in the game is a barrier. During play, players also share tips and context with each other: "*If you hear water, walk away* [to avoid pitfalls]".

When playing with strangers, some announce their disability to both contextualize their actions and to ask for certain behaviors from their team members. In WoW, C11 uses the chat to inform being unable to read messages and requiring someone from the party to be close so C11 can follow their avatar. C1 highlights that playing with others usually goes well when co-players are communicative, giving quick indications for navigation ("*just say up, down, left, right*"). C1 also mentions the use of quick dialogue inputs in *Resident Evil* (e.g., pressing a button combination makes the character say "*Follow me*") that also allows aligning the camera with the player speaking. In some games (e.g., L4D2, Diablo), players become 'idle' to follow other players automatically.

*4.7.2 Engaging with spectators & audience.* Friend spectators (co-located or remote through audio call) and live audiences (stream chat) take an active role in explaining controls and menus, reading onscreen prompts, and helping with navigation and aiming, which is often fundamental to proceed in the game: "*If I can get a handle on the gameplay, and if you guys are able to help out with inventory and stuff, I'm pretty sure I'll play a lot more*" (C1, playing *Dark Souls*). At various moments, they have to clarify what, why, and how something is happening. As aforementioned, C4 was unable to interact with an NPC in *Stardew* and assumed there was no one there and left. Immediately, the live audience described the event in the chat, allowing C4 to understand what had actually happened. Often, spectators give indications that guide navigation: "*You have to go away from the screaming [sound]*" (audience to C1, when playing *Destiny*), which can become part of the core gameplay loop.

*Cumbersome assistance.* There is often a **delay** in the stream, which, coupled with the multiple audience members trying to assist, can lead to difficult interactions. Spectators also have no way to indicate **accurate distances or degrees** (e.g., "*a bit to the left*"). Even when spectators are co-present, there are limits to their usefulness in time-sensitive events. Importantly, players have to forgo some agency for audiences to be able to help. While playing *Destiny*, the stream chat asks C1 to be still for a moment so they can give step-by-step indications. C1 adheres but says "*Man, I'm so impatient*".

*4.7.3 Sharing control with co-pilot.* In some videos, players were sharing control with a partner (Co). This happened in all playthroughs from C3 (remotely, through Parsec) and C8 (co-located, using the co-pilot feature that lets two controllers act as one). This practice can potentially transform any game into a cooperative experience where inaccessible content is either played or partially assisted by sighted players. Despite theoretically lowering player agency, for C3, the collaboration actually improves the experience: "*I love the fact we can work as a team on this*" (playing *Horizon*). Players work together on different challenges, taking on different roles according to their abilities. For example, when playing *Horizon*, Co3 first describes the scene: "*We are on a rock high up, we're above the enemy camp*". Then, players usually discuss and strategize on how to approach (i.e., silently approaching or "*sniper*" approach). Players collaborate to tag enemies, with Co3 aiming and prompting C3 to tag. When shooting, C3 draws the arrow and waits for Co3's cue ("*tell me when*"). This example illustrates the rich experience and collaboration that exists when playing with co-pilot.

*Agency shift.* Players share control of the experience in different ways. C3 was responsible for combat sections (with verbal guidance from Co3), while Co3 navigated between spaces and helped during visual puzzles. Co3 also spontaneously prompts C3 to do simple tasks (e.g., jump) or trigger interactions. A similar distribution of responsibilities is agreed between C8 and Co8, with players at times not fully aware of how to perform one another's: [Co8 prompts C8 to jump when playing Neverwinter]; "You can jump too!" (C8); "I don't remember how to jump [laughs]" (Co8).

Sudden shifts in control between players were also observed. For instance, when playing *Horizon*, Co3 is navigating unaware of an enemy to their left (out of the camera view). C3 hears it, warns the partner, and turns the camera in the enemy's direction: "*Hang on, watcher*". Until that point, the camera was exclusively controlled by the sighted player: "*That was a really good example, though, of agency, because I was able to turn the camera while you were moving, and *basically just show you what I was hearing*". Still, shifts of control emerged mostly from sighted players, when dealing with critical situations (e.g., in *Horizon*, Co3 takes control to flee when it becomes severely damaged in combat).

## 5 DISCUSSION

Blind people are (with a certain degree of success) playing vision-centric mainstream games despite the myriad of accessibility problems on top of the intended game challenge. They have to be ingenious beyond the intended game design to come up with solutions that work for them. We have illustrated throughout the findings a multitude of barriers (RQ1) associated with 1) perceiving the surroundings, 2) wayfinding, 3) perspective-based gameplay, 4) interaction patterns, 5) cognitive load; and discussed the coping strategies associated with them (RQ2). Furthermore, we described 6) how automation/customization features are impacting the gameplay for the sake of accessibility; and 7) how blind players are sharing play with others (RQ2). We extend previous work on characterizing blind people's experiences with digital games [2, 4, 18, 38, 45, 52], providing a deeper understanding of the actual experience when playing mainstream games. Our results portray how game design is currently failing to meet blind players' needs (e.g., perception of occlusions) and call forth new approaches where their preferred strategies to navigate and interact (e.g., save states) are considered and integrated. Below, we summarize and discuss these implications for inclusive game design.

### 5.1 Orientation and mobility strategies in game environments

While sighted people can quickly grasp the scene visually, blind players have to **bump and act upon the environment to understand it**. Through repetition, players form a mental map based on unique audio **landmarks representing locations**. When these landmarks naturally occur in the design (e.g., torches on both sides of a door), navigation becomes an enjoyable part of the experience. Further, players appropriate and create new landmarks, using items to mark the environment (e.g., portals), leaving enemies behind to re-identify rooms, or following along with NPCs.

The strategies employed to independently explore and map environments constitute a challenge beyond those intended by design—one that, as observed, can be part of the fun. However, the increased effort sets a higher bar to entry for any potential blind player. Also, they can result in frustrating experiences, with examples of players getting disoriented and having to either backtrack to a familiar location or feel around the scene again. These experiences resemble the challenges faced in the real world where blind people often avoid visiting unfamiliar places by themselves due to a lack of knowledge about the environment [17, 33, 49]. In the future, designers should consider how to support and integrate blind people's current orientation and mobility strategies (e.g., the use of landmarks or an initial description of the environment) from the ground up to optimize the initial exploration of an area.

*5.1.1 Egocentric vs Allocentric spatial representations.* Spatial representations of (real or virtual) environments can take an egocentric or allocentric perspective. Egocentric defines distances and directions relative to oneself akin to what is provided by games with a first- or third-person perspective. An egocentric perspective is strongly tied to how blind people navigate unfamiliar and complex places and how they tend to form route-based sequential representations of such environments [32, 42, 44].

We observed this behavior frequently as players relied on a list of specific steps that include landmarks, turns, and distances in order to navigate and progress in the game. While landmarks play an essential role in this process, **camera handling posed many challenges** that influenced perspective perception resulting in confusion or disorientation. This was especially problematic because route-based knowledge does not provide a holistic view of the environment making it harder to recover from errors.

An allocentric perspective, on the other hand, defines the locations of relevant elements relative to each other and to other external references, in what is usually linked with a more holistic, map-like representation of the environment [9, 32]. A parallel can be drawn to a top-down perspective where movement is based on a fixed axis and sound direction is absolute (e.g., sound from the left is always from the left-end side of the map).

Our observations and players' comments suggest that navigating in games with a top-down perspective resulted in more reliable mental models of the environment. However, in gaming contexts, it is not clear what are the consequences of these types of navigation and perspective. Future work is needed to further explore how they fit into blind people's needs and preferences, and if providing both (or combined) options could improve accessibility.

### 5.2 Audio design and game mechanics as assistive technologies

Blind people are able to play vision-centric games when there is appropriate audio feedback. **Spatial sound and soundscape changes** (e.g., ambient, footstep sounds) play a fundamental role in creating an accessible environment. These elements should distinguish objects, areas, and their characteristics (e.g., open or indoor location). Audio cues and landmarks are used by blind people when navigating the real-world [26, 48] and therefore are also pivotal in approaches that use virtual environments for training navigation skills [12, 14, 27]. In contrast, **game design is offering limited sets of unique sounds** and often disregards their meaning and intuitiveness for someone who solely relies on audio. This adds to the cognitive load and demands more memorization from players.

Some design decisions may seem minor, but have a significant impact on accessibility. For instance, conveying movement sounds when the player is actually running against an obstacle impedes players from having a clear perception of collisions. Having overwhelming sounds (e.g., a loud generator) over important information, or enticing for irrelevant elements happens recurrently. It is up to designers to create environments that have as much attention to detail for auditory feedback as visual. Additionally, even though haptics are used by some titles to complement information (e.g., vibration when the avatar takes damage) and augment the perception over fine movement (e.g., vibration intensity to convey distance to a target), they should be further explored in gaming.

**Player onboarding is particularly neglected**—learnability patterns that help players familiarize themselves with the gameplay are almost exclusively visual [37]. As a consequence, players usually resort to external sources (e.g., walkthroughs) and other people to learn how to play. In order to ensure a smooth learning curve for blind players, we need to explore new ways to introduce mechanics and provide contextual hints through audio.

*5.2.1 Appropriation of game elements.* Even for games that do not present any accessibility whatsoever, players take it into their own hands, 'hacking' their way through the gameplay. Players find inventive ways of **leveraging game features** and elements to understand, navigate and interact with the environment. From using objects as virtual canes in order to perceive collisions, materials, and distance, to landscaping the environment (e.g., placing bushes and torches) to orient themselves. Other helpful features include teleportation and save states. Unlike in the real world, players have a panoply of methods to successfully (and enjoyably) navigate in games. A potential avenue is to provide tools enabling the integration of these sorts of elements in the experience (e.g., the ability to mark the environment with beacons or signs). Besides, we should think of ways to integrate such elements, by respecting the context of the game, ensuring they still feel "*natural*" (in the words of C9, like surrounding a patio with bushes to landmark it).

### 5.3 Difficulty in perceiving the bigger picture

Games usually offer a rich visual interface that simultaneously conveys various details about the environment and game state. When navigating, vision allows to discern the different pathways that can be taken, from a distance. When in combat, it allows for an accurate estimation of enemies' direction, proximity, and behavior. Such is very difficult for blind players, as they have to first resort to 'feeling around' the scene and create a mental map of it.

This results in **higher cognitive demand**, with players keeping track of many things at once and getting confused about their whereabouts and game state. Currently, navigation in games (even considering semi-automatic approaches) does not provide the opportunity for blind players to explore what they want. They are often unaware of interactive objects and entities, and do not have all information to best deal with them (e.g., not having a "tone change" for when an enemy looks away, or to indicate occlusions when shooting or taking cover). There is a need to **find ways to convey concurrent information** through sound and haptics, that can still be intuitively associated with its meaning and ensure players take informed decisions.

### 5.4 Spectrum between agency and full accessibility (automation)

Most accessibility features automate parts of the experience or diminish the challenge (e.g., invisible when prone). In some cases, they totally alter the challenge (e.g., auto-steer in a racing game). When playing with others, the gameplay is broken up into smaller challenges and shared with co-players, spectators, and/or co-pilots. All these **reduce the level of interactivity offered by the game**, which can ultimately diminish player agency [11]. At stake is the balance between ensuring accessibility and still offering meaningful decisions and challenging interactions [18, 41].

Still, it should be acknowledged that players might enjoy a game in a more passive way, like C10 playing Forza, simply pressing an input to advance while enjoying the soundscape. In some narrative-oriented games, one is solely required to move in the world to advance the story (popularly called 'walking simulators'). Such games present a low level of interactivity, but, as with any game, still empower some form of agency. Previous work [11] suggests that agency should be framed in video games not only as objective actions and outcomes inputted by the player but also as the perceived impact and reflections these actions have in the narrative.

While player agency is a central factor of digital games, the extent to which one gets involved with the experience will highly depend on individual preferences. Some players are primarily interested in discovering and sharing the game's content with others, instead of having control and action over the gameplay [6]. The principle should be inclusion, even if in different ways. Blind players will invariably experience games differently from their sighted peers, but they must have the opportunity to "*join*" and see it as "*normal*" (C9). In addition to trying to provide equal access, we can focus on exploring, improving, and contextualizing the changes needed to welcome blind players (e.g., playing *Horizon* in co-pilot is essentially transforming a single-player game into a collaborative experience).

*5.4.1 Agency in co-play.* Co-play experiences were fruitful, with players able to play games that were mostly unplayable, with **spectators and co-players assisting in some tasks** (e.g., navigation between points of interest, visual puzzles). In some cases, their role was just to explain or demonstrate how the game works. When playing the game for the first time, blind players would experiment and take on lesser roles while the collaboration slowly evolved and morphed into a split between the two players. Co-pilot features allow blind and sighted players to experience games together in a different way than before. We should think of how game design and new technologies can support these types of experiences, addressing cumbersome assistance (e.g., inaccurate indications, delay) and maximizing enjoyment for all parties. While the fun can also come from exploring how to split controls, considering this possibility right from the start might create new engaging experiences.

### 5.5 Playing your own game and curating accessibility pathways

The efforts behind playing a mainstream game often include the work of picking out the most accessible experience the game can offer to someone blind. Players methodically explore different options, base their choices on recommendations of others (e.g., online videos, guides) or even rely on community-created *mods*. These options include enabling particular settings, using particularly useful characters and abilities, and playing in the most accessible maps and levels. *Mods* can be a vehicle to accessibility for the most popular games where natural big modding communities are created. Although none of these are a replacement for inclusive design, providing **ways for the community to shape and extend the experience of a game** can bridge gaps in accessibility that were not accounted for during development.

The search for the **optimal accessibility pathway** mimics the common behavior of players sharing and looking for the top-performing builds (i.e. the arrangement of stat points or gear that allows players to achieve the best results for each play style) and optimal paths, but with a completely different focus. While it can somewhat narrow players' choices, finding these paths is akin to discovering and optimizing player characters which can be part of the game's appeal. It remains to be seen whether this actually improves or impedes the experience of its players (even if at the cost of accessibility). Also, being dependent on these features, blind players end up being excluded in modes where they are not available or allowed (in particular competitive multiplayer).

We need to not only look for ways to improve the design from the start (e.g., designing specific roles and abilities that cater to blind players), but also look back at ways to enable these strategies in previously released games. Some game developer companies make **guides available for new and old games**, explicitly made for blind and visually impaired players [39]. Without these efforts, we face the risk of perpetually having hundreds of thousands of past games available to sighted users simply being inaccessible to blind players (or requiring extraordinary commitment from the player).

### 5.6 Limitations

Online ethnographic methods apply traditional qualitative research tools to digital settings [35]. While they provide an opportunity to gather rich contextual portrayals of lived experiences in an unobtrusive and reliable way, they come with limitations [35]. Importantly, although we ensured a careful sampling, our results portray the experiences of (mostly) expert gamers who, on top of enduring significant efforts to play mainstream games, share them on YouTube. They do not include potentially relevant experiences that could be collected through offline methods (e.g., in-person interviews). Also, this work purposefully does not reflect the experiences of a general population, given that, as shown by previous work, most blind people resort to accessible alternatives (audio games) [2, 18, 45, 52].

While our analysis captures a variety of genres and types of navigation, some are absent. In particular, 'side-scrolling' games comprise a common type of navigation, but there were no games of this genre in any of the channels included (apart from fighting games, which were excluded, given they do not present navigation challenges). Lastly, it is worth pointing out that haptic feedback was not observable through the videos, and was only detected when the player in the video specifically mentioned it.

## 6  CONCLUSION

Inclusive design is crucial and more so as gaming has become one of the most popular entertainment mediums and continues to grow. In this article, we have described the recorded experiences of multiple blind content creators playing mainstream games, with a particular focus on the strategies that were implemented to overcome their lack of accessibility. We described how blind players leverage specific game features and mechanics to simplify or automate tasks, and to receive additional audible information. Some are able to create their own ways of playing, overcoming many of the accessibility barriers the game presents and sometimes becoming fairly proficient. These strategies show potential new ways for designing games to be accessible while maintaining the core gameplay experience.

## ACKNOWLEDGMENTS

We thank all the content-creators for sharing their experiences, and especially those who responded to our contacts with interest, enthusiasm, and availability to further discuss the work. This work was supported by FCT through project "Plug n' Play: Exploring Asymmetry and Modularity for Inclusive Game Design" ref. 2022.08895.PTDC, project "Virtual Reality Accessible to Visually Impaired People" ref. 2022.08286.PTDC, scholarships ref. UI/BD/151178/2021 and ref. 2022.12448.BD, and the LASIGE Research Unit, ref. UIDB/00408/2020 and ref. UIDP/00408/2020.

## REFERENCES

[1] Ronny Andrade, Steven Baker, Jenny Waycott, and Frank Vetere. 2018. Echo-House: Exploring a Virtual Environment by Using Echolocation. In *Proceedings of the 30th Australian Conference on Computer-Human Interaction* (Melbourne, Australia) *(OzCHI '18)*. Association for Computing Machinery, New York, NY, USA, 278–289. https://doi.org/10.1145/3292147.3292163

[2] Ronny Andrade, Melissa J. Rogerson, Jenny Waycott, Steven Baker, and Frank Vetere. 2019. Playing Blind: Revealing the World of Gamers with Visual Impairment. In *Proceedings of the 2019 CHI Conference on Human Factors in Computing Systems* (Glasgow, Scotland Uk) *(CHI '19)*. Association for Computing Machinery, New York, NY, USA, 1–14. https://doi.org/10.1145/3290605.3300346

[3] Lisa Anthony, YooJin Kim, and Leah Findlater. 2013. Analyzing User-Generated Youtube Videos to Understand Touchscreen Use by People with Motor Impairments. In *Proceedings of the SIGCHI Conference on Human Factors in Computing Systems* (Paris, France) *(CHI '13)*. Association for Computing Machinery, New York, NY, USA, 1223–1232. https://doi.org/10.1145/2470654.2466158

[4] Dominique Archambault, Roland Ossmann, Thomas Gaudy, and Klaus Miesenberger. 2007. Computer games and visually impaired people. *Upgrade* (Jan 2007).

[5] Matthew T. Atkinson, Sabahattin Gucukoglu, Colin H. C. Machin, and Adrian E. Lawrence. 2006. Making the Mainstream Accessible: Redefining the Game. In *Proceedings of the 2006 ACM SIGGRAPH Symposium on Videogames* (Boston, Massachusetts) *(Sandbox '06)*. Association for Computing Machinery, New York, NY, USA, 21–28. https://doi.org/10.1145/1183316.1183321

[6] Richard Bartle. 1996. Hearts, clubs, diamonds, spades: Players who suit MUDs. *Journal of MUD Research* (06 1996).

[7] Brian Bors. 2015. The current state of game accessibility guidelines. https://www.game-accessibility.com/documentation/accessibility-guidelines/ Library Catalog: www.game-accessibility.com.

[8] Virginia Braun and Victoria Clarke. 2006. Using thematic analysis in psychology. *Qualitative Research in Psychology* 3, 2 (Jan 2006), 77–101. https://doi.org/10.1191/1478088706qp063oa

[9] Anke Brock. 2013. *Interactive maps for visually impaired people: design, usability and spatial cognition*. Ph. D. Dissertation. Université Toulouse 3 Paul Sabatier.

[10] David Byrne. 2022. A worked example of Braun and Clarke's approach to reflexive thematic analysis. *Quality & Quantity* 56, 3 (Jun 2022), 1391–1412. https://doi.org/10.1007/s11135-021-01182-y

[11] Tom Cole and Marco Gillies. 2021. Thinking and Doing: Challenge, Agency, and the Eudaimonic Experience in Video Games. *Games and Culture* 16, 2 (Mar 2021), 187–207. https://doi.org/10.1177/1555412019881536

[12] Erin C Connors, Elizabeth R Chrastil, Jaime Sánchez, and Lotfi B Merabet. 2014. Virtual environments for the transfer of navigation skills in the blind: a comparison of directed instruction vs. video game based learning approaches. *Frontiers in human neuroscience* 8 (2014), 223.

[13] Jared Duval, Ferran Altarriba Bertran, Siying Chen, Melissa Chu, Divya Subramonian, Austin Wang, Geoffrey Xiang, Sri Kurniawan, and Katherine Isbister. 2021. Chasing Play on TikTok from Populations with Disabilities to Inspire Playful and Inclusive Technology Design. In *Proceedings of the 2021 CHI Conference on Human Factors in Computing Systems* (Yokohama, Japan) *(CHI '21)*. Association for Computing Machinery, New York, NY, USA, Article 492, 15 pages. https://doi.org/10.1145/3411764.3445303

[14] Agebson Rocha Façanha, Ticianne Darin, Windson Viana, and Jaime Sánchez. 2020. O&M Indoor Virtual Environments for People Who Are Blind: A Systematic Literature Review. *ACM Trans. Access. Comput.* 13, 2, Article 9a (aug 2020), 42 pages. https://doi.org/10.1145/3395769

[15] IGDA GASIG. 2020. The Accessibility in Last of Us Part II: A 3 Year Journey. https://youtu.be/5HDdino-umA?t=350

[16] Kathrin Gerling and Laura Buttrick. 2014. Last Tank Rolling: Exploring Shared Motion-Based Play to Empower Persons Using Wheelchairs. In *Proceedings of the First ACM SIGCHI Annual Symposium on Computer-Human Interaction in Play* (Toronto, Ontario, Canada) *(CHI PLAY '14)*. Association for Computing Machinery, New York, NY, USA, 415–416. https://doi.org/10.1145/2658537.2661303

[17] Nicholas A Giudice. 2018. Navigating without vision: Principles of blind spatial cognition. In *Handbook of behavioral and cognitive geography*. Edward Elgar Publishing.

[18] David Gonçalves, André Rodrigues, and Tiago Guerreiro. 2020. Playing With Others: Depicting Multiplayer Gaming Experiences of People With Visual Impairments. In *The 22nd International ACM SIGACCESS Conference on Computers and Accessibility* (Virtual Event, Greece) *(ASSETS '20)*. Association for Computing Machinery, New York, NY, USA, Article 22, 12 pages. https://doi.org/10.1145/3373625.3418304

[19] David Gonçalves, André Rodrigues, Mike L. Richardson, Alexandra A. de Sousa, Michael J. Proulx, and Tiago Guerreiro. 2021. Exploring Asymmetric Roles in Mixed-Ability Gaming. In *Proceedings of the 2021 CHI Conference on Human Factors in Computing Systems* (Yokohama, Japan) *(CHI '21)*. Association for Computing Machinery, New York, NY, USA, Article 114, 14 pages. https://doi.org/10.1145/3411764.3445494

[20] Dimitris Grammenos. 2008. Game over: learning by dying. In *Proceedings of the SIGCHI Conference on Human Factors in Computing Systems (CHI '08)*. Association for Computing Machinery, 1443–1452. https://doi.org/10.1145/1357054.1357281

[21] Dimitris Grammenos, Anthony Savidis, and Constantine Stephanidis. 2009. Designing universally accessible games. *Computers in Entertainment* 7, 1 (Feb 2009), 8:1–8:29. https://doi.org/10.1145/1486508.1486516

[22] D Wesley Grantham, Benjamin WY Hornsby, and Eric A Erpenbeck. 2003. Auditory spatial resolution in horizontal, vertical, and diagonal planes. *The Journal of the Acoustical Society of America* 114, 2 (2003), 1009–1022.

[23] João Guerreiro, Daisuke Sato, Dragan Ahmetovic, Eshed Ohn-Bar, Kris M. Kitani, and Chieko Asakawa. 2020. Virtual navigation for blind people: Transferring route knowledge to the real-World. *International Journal of Human-Computer Studies* 135 (Mar 2020), 102369. https://doi.org/10.1016/j.ijhcs.2019.102369

[24] João Guerreiro, Dragan Ahmetovic, Kris M. Kitani, and Chieko Asakawa. 2017. Virtual Navigation for Blind People: Building Sequential Representations of the Real-World. In *Proceedings of the 19th International ACM SIGACCESS Conference on Computers and Accessibility* (Baltimore, Maryland, USA) *(ASSETS '17)*. Association for Computing Machinery, New York, NY, USA, 280–289. https://doi.org/10.1145/3132525.3132545

[25] Ian Hamilton, Barrie Ellis, Gareth Ford-Williams, Lynsey Graham, Dimitris Grammenos, Headstrong Games, Ed Lee, Jake Manion, and Thomas Westin. 2012. Game accessibility guidelines | A straightforward reference for inclusive game design. http://gameaccessibilityguidelines.com/

[26] Athanasios Koutsoklenis and Konstantinos Papadopoulos. 2011. Auditory cues used for wayfinding in urban environments by individuals with visual impairments. *Journal of Visual Impairment & Blindness* 105, 10 (2011), 703–714.

[27] Orly Lahav and David Mioduser. 2008. Construction of cognitive maps of unknown spaces using a multi-sensory virtual environment for people who are blind. *Computers in Human Behavior* 24, 3 (2008), 1139–1155.

[28] Franklin Mingzhe Li, Jamie Dorst, Peter Cederberg, and Patrick Carrington. 2021. Non-Visual Cooking: Exploring Practices and Challenges of Meal Preparation by People with Visual Impairments. In *The 23rd International ACM SIGACCESS Conference on Computers and Accessibility* (Virtual Event, USA) *(ASSETS '21)*. Association for Computing Machinery, New York, NY, USA, Article 30, 11 pages. https://doi.org/10.1145/3441852.3471215

[29] Franklin Mingzhe Li, Franchesca Spektor, Meng Xia, Mina Huh, Peter Cederberg, Yuqi Gong, Kristen Shinohara, and Patrick Carrington. 2022. "It Feels Like Taking a Gamble": Exploring Perceptions, Practices, and Challenges of Using Makeup and Cosmetics for People with Visual Impairments. In *Proceedings of the 2022 CHI Conference on Human Factors in Computing Systems* (New Orleans, LA, USA) *(CHI '22)*. Association for Computing Machinery, New York, NY, USA, Article 266, 15 pages. https://doi.org/10.1145/3491102.3517490

[30] Masaki Matsuo, Takahiro Miura, Masatsugu Sakajiri, Junji Onishi, and Tsukasa Ono. 2016. ShadowRine: Accessible game for blind users, and accessible action RPG for visually impaired gamers. In *2016 IEEE International Conference on Systems, Man, and Cybernetics (SMC)*. 002826–002827. https://doi.org/10.1109/SMC.2016.7844667

[31] Microsoft. 2022. Xbox Accessibility Guidelines - Microsoft Game Dev. https://docs.microsoft.com/en-us/gaming/accessibility/guidelines

[32] Susanna Millar. 1994. *Understanding and Representing Space: Theory and Evidence from Studies with Blind and Sighted Children*. Oxford University Press. https://doi.org/10.1093/acprof:oso/9780198521426.001.0001

[33] Karin Müller, Christin Engel, Claudia Loitsch, Rainer Stiefelhagen, and Gerhard Weber. 2022. Traveling more Independently: A Study on the Diverse Needs and Challenges of People with Visual or Mobility Impairments in Unfamiliar Indoor Environments. *ACM Transactions on Accessible Computing (TACCESS)* 15, 2 (2022), 1–44.

[34] Vishnu Nair, Jay L Karp, Samuel Silverman, Mohar Kalra, Hollis Lehv, Faizan Jamil, and Brian A. Smith. 2021. NavStick: Making Video Games Blind-Accessible via the Ability to Look Around. In *The 34th Annual ACM Symposium on User Interface Software and Technology* (Virtual Event, USA) *(UIST '21)*. Association for Computing Machinery, New York, NY, USA, 538–551. https://doi.org/10.1145/3472749.3474768

[35] Thaysa Nascimento, Maribel Carvalho Suarez, and Roberta Dias Campos. 2022. An integrative review on online ethnography methods: differentiating theoretical bases, potentialities and limitations. *Qualitative Market Research: An International Journal* 25, 4 (Jan 2022), 492–510. https://doi.org/10.1108/QMR-07-2021-0086

[36] Naughty Dog. 2020. *The Last of Us Part II*. Game [PlayStation 4]. San Mateo, California, United States, Sony Interactive Entertainment..

[37] Lev Poretski and Anthony Tang. 2022. Press A to Jump: Design Strategies for Video Game Learnability. In *Proceedings of the 2022 CHI Conference on Human Factors in Computing Systems* (New Orleans, LA, USA) *(CHI '22)*. Association for Computing Machinery, New York, NY, USA, Article 155, 26 pages. https://doi.org/10.1145/3491102.3517685

[38] John R. Porter and Julie A. Kientz. 2013. An Empirical Study of Issues and Barriers to Mainstream Video Game Accessibility. In *Proceedings of the 15th International ACM SIGACCESS Conference on Computers and Accessibility* (Bellevue, Washington) *(ASSETS '13)*. Association for Computing Machinery, New York, NY, USA, Article 3, 8 pages. https://doi.org/10.1145/2513383.2513444

[39] Nelson Régo. 2018. EA Sports Launches Accessibility Portal for Blind Gamers. https://coolblindtech.com/ea-sports-launches-accessibility-portal-for-blind-gamers/

[40] Alexa F. Siu, Mike Sinclair, Robert Kovacs, Eyal Ofek, Christian Holz, and Edward Cutrell. 2020. Virtual Reality Without Vision: A Haptic and Auditory White Cane to Navigate Complex Virtual Worlds. In *Proceedings of the 2020 CHI Conference on Human Factors in Computing Systems* (Honolulu, HI, USA) *(CHI '20)*. Association for Computing Machinery, New York, NY, USA, 1–13. https://doi.org/10.1145/3313831.3376353

[41] Brian A. Smith and Shree K. Nayar. 2018. The RAD: Making Racing Games Equivalently Accessible to People Who Are Blind. In *Proceedings of the 2018 CHI Conference on Human Factors in Computing Systems* (Montreal QC, Canada) *(CHI '18)*. Association for Computing Machinery, New York, NY, USA, 1–12. https://doi.org/10.1145/3173574.3174090

[42] Catherine Thinus-Blanc and Florence Gaunet. 1997. Representation of space in blind persons: vision as a spatial sense? *Psychological bulletin* 121, 1 (1997), 20.

[43] Shari Trewin, Vicki L. Hanson, Mark R. Laff, and Anna Cavender. 2008. PowerUp: An Accessible Virtual World. In *Proceedings of the 10th International ACM SIGACCESS Conference on Computers and Accessibility* (Halifax, Nova Scotia, Canada) *(Assets '08)*. Association for Computing Machinery, New York, NY, USA, 177–184. https://doi.org/10.1145/1414471.1414504

[44] Simon Ungar. 2018. Cognitive mapping without visual experience. In *Cognitive Mapping*. Routledge, 221–248.

[45] Michael Urbanek and Florian Güldenpfennig. 2019. Unpacking the Audio Game Experience: Lessons Learned from Game Veterans. In *Proceedings of the Annual Symposium on Computer-Human Interaction in Play* (Barcelona, Spain) *(CHI PLAY '19)*. Association for Computing Machinery, New York, NY, USA, 253–264. https://doi.org/10.1145/3311350.3347182

[46] Elizabeth M Wenzel, Marianne Arruda, Doris J Kistler, and Frederic L Wightman. 1993. Localization using nonindividualized head-related transfer functions. *The Journal of the Acoustical Society of America* 94, 1 (1993), 111–123.

[47] T. Westin. 2004. Game accessibility case study: Terraformers – a real-time 3d graphic game. In *In Proc. of the The Fifth International Conference on Disability, Virtual Reality and Associated Technologies*. 95–100.

[48] William R Wiener, Richard L Welsh, and Bruce B Blasch. 2010. *Foundations of orientation and mobility*. Vol. 1. American Foundation for the Blind.

[49] Michele A Williams, Amy Hurst, and Shaun K Kane. 2013. " Pray before you step out" describing personal and situational blind navigation behaviors. In *Proceedings of the 15th International ACM SIGACCESS Conference on Computers and Accessibility*. 1–8.

[50] Jingyi Xie, Na Li, Sooyeon Lee, and John M. Carroll. 2022. YouTube Videos as Data: Seeing Daily Challenges for People with Visual Impairments During COVID-19. In *Proceedings of the 2022 ACM Conference on Information Technology for Social Good* (Limassol, Cyprus) *(GoodIT '22)*. Association for Computing Machinery, New York, NY, USA, 218–224. https://doi.org/10.1145/3524458.3547224

[51] Bei Yuan and Eelke Folmer. 2008. Blind Hero: Enabling Guitar Hero for the Visually Impaired. In *Proceedings of the 10th International ACM SIGACCESS Conference on Computers and Accessibility* (Halifax, Nova Scotia, Canada) *(Assets '08)*. Association for Computing Machinery, New York, NY, USA, 169–176. https://doi.org/10.1145/1414471.1414503

[52] Bei Yuan, Eelke Folmer, and Frederick C. Harris. 2011. Game Accessibility: A Survey. *Univers. Access Inf. Soc.* 10, 1 (mar 2011), 81–100. https://doi.org/10.1007/s10209-010-0189-5

[53] Yuhang Zhao, Cynthia L. Bennett, Hrvoje Benko, Edward Cutrell, Christian Holz, Meredith Ringel Morris, and Mike Sinclair. 2018. Enabling People with Visual Impairments to Navigate Virtual Reality with a Haptic and Auditory Cane Simulation. In *Proceedings of the 2018 CHI Conference on Human Factors in Computing Systems* (Montreal QC, Canada) *(CHI '18)*. Association for Computing Machinery, New York, NY, USA, 1–14. https://doi.org/10.1145/3173574.3173690